# INSTANTLY DEPLOYABLE EXPERT KNOWLEDGE – NETWORKS OF KNOWLEDGE ENGINES

Version 1.0
2018-11-07

**Bernhard Bergmair\*, Thomas Buchegger\*, Johann Hoffelner\*, Gerald Schatz\*, Siegfried Silber\*, Johannes Klinglmayr\***

\* Linz Center of Mechatronics GmbH

**Corresponding authors:**
Johannes Klinglmayr (johannes.klinglmayr@lcm.at),
Bernhard Bergmair (bernhard.bergmair@lcm.at)

## ABSTRACT

*Knowledge and information are becoming the primary resources of the emerging information society. To exploit the potential of available expert knowledge, comprehension and application skills (i.e. expert competences) are necessary. The ability to acquire these skills is limited for any individual human. Consequently, the capacities to solve problems based on human knowledge in a manual (i.e. mental) way are strongly limited.*

*We envision a new systemic approach to enable scalable knowledge deployment without expert competences. Eventually, the system is meant to instantly deploy humanity's total knowledge in full depth for every individual challenge. To this end, we propose a socio-technical framework that transforms expert knowledge into a solution creation system. Knowledge is represented by automated algorithms (knowledge engines). Executable compositions of knowledge engines (networks of knowledge engines) generate requested individual information at runtime. We outline how these knowledge representations could yield legal, ethical and social challenges and nurture new business and remuneration models on knowledge. We identify major technological and economic concepts that are already pushing the boundaries in knowledge utilisation: E.g. in artificial intelligence, knowledge bases, ontologies, advanced search tools, automation of knowledge work, the API economy. We indicate impacts on society, economy and labour. Existing developments are linked, including a specific use case in engineering design.*

# 1 INTRODUCTION

For decades we experience an ongoing structural shift in value creation: from agricultural and industrial production to services and, more recently, to information- and knowledge-based services.[1] **Information and knowledge are becoming primary resources of the emerging knowledge society.[2]**

When facing any kind of problem (e.g. repairing the lawn mower, or fixing software configuration issues) we got used to search the internet (or other databases) for information on existing solutions. But existing solutions are of limited use, to answer individual questions that have never been asked before, or to foster innovation by solving unprecedented problems. In such cases knowledge must be applied to generate accurate solutions upon request. Within this context knowledge is related to the capacity of solving a class of problems.[3]

For a human, to comprehend knowledge to the degree that they can apply it to solve problems might be desirable and satisfying – but it is tedious. Thus, the number of problems that can be solved by one person based on self-gained knowledge is limited.[1]

One widespread alternative to gaining knowledge oneself is to consult someone with extended knowledge access – an expert. This has led to sophisticated forms of differentiation of labour and finally to a flourishing industry of knowledge-intensive services. But again, one encounters strong limitations: The findability and availability of experts. Finding the right expert necessitates knowledge about which experts can solve which classes of problems. And even if one has identified matching experts, experts are scarce and might be needed by many. Thus, services of experts are expensive and are (in the non-public domain) merely employed for creating solutions with expected financial payback. Consequently, the capacities to solve problems based on human knowledge in a manual (i.e. mental) way are very limited.

# 2 NETWORKS OF KNOWLEDGE ENGINES

## 2.1 VISION

We envision a new culture and technology framework to enable scalable knowledge utilisation for solving human problems beyond those restrictions. **In its limit, the envisioned framework enables everyone to utilise humanity's total knowledge in full depth for each individual challenge.** While it is of course utopian to expect the full realisation of this vision any time soon, it can provide the course for a self-determined humankind in the beginning age of artificial intelligence.

This paper proposes a taxonomy to describe the vision, its elements and necessary framework conditions. It provides an interdisciplinary overview of ongoing initiatives, that are already contributing to this vision, including examples of first operational elements.

## 2.2 THE NOTION(S) OF KNOWLEDGE

Speaking about interdisciplinarity: The notion of 'knowledge' lacks a common definition that could be agreed on by all relevant disciplines. For the context of this paper we consider the following features of knowledge decisive:[4]

---

[1] If the utilisation of a piece of information necessitates no deep comprehension, e.g. following the instructions of a recipe for pancakes, then the information does not represent the knowledge to solve a class of problems, but rather is the solution to the specific problem "how to make a pancake".

**Knowledge has a practical aspect.** Knowledge does not only consist in knowing objective facts. Knowledge helps to solve problems – not one specific problem, but a whole class of problems. And "actionable knowledge" might become the central resource of economies.[5]

**Knowledge is person-bound or (externally) represented.** Knowledge representations can comprise facts and algorithms.

**Knowledge has a normative structure.** Knowledge consists of claims (e.g. claiming to be a solution for a class of problems). The claims need to be recognised by others as "successful" (e.g. successful solutions).

**Knowledge is internally and externally networked.** It consists of an internal network (e.g. a logical structure) of elements that are considered knowledge themselves. And knowledge is necessarily linked to externally existing knowledge.

**Knowledge is dynamic.** "Knowledge is acquired and disposed, is recognised, used/applied, sold and bought, written down, transferred, shared or kept secret, reformulated, etc. … Knowledge can also be forgotten or disappear when unused for a long time."

These features of knowledge along with the subsequent description of the proposed vision support an important hypothesis: A framework to utilise humankind's total knowledge naturally exhibits intrinsic features of knowledge. The framework is self-similar: While operating with knowledge, the framework itself shows the features of knowledge. Such a framework could constitute or enable a decentralised human **"collective intelligence"**.[6] However, what is not yet described by these features, is the scalability and ease of knowledge utilisation.

## 2.3 AUTOMATED APPLICATION OF KNOWLEDGE

A culture and technology framework to enable scalable knowledge utilisation must make use of automated application of knowledge. While in the last decades the automation of production has been advanced to a very high level, automation of knowledge work seemed impossible – except for minor isolated applications. But due to advances in artificial intelligence – understood as a wide field of diverse approaches – more and more knowledge can be depicted by software. Prominent developments comprise machine learning, expert systems, automatic engineering design processes[7], model-based systems engineering, simulation and optimisation, social physics, ontologies and semantic reasoning. In all these domains, experts are advancing the automation of specific knowledge-intensive tasks (i.e. problem solving). We expect many of these problem solvers to be offered as automated services. The workload capacity of these automated services is extremely high and marginal costs will approach zero. Hence, it might even be sufficient (and favoured by markets) to have a single provider of a specific expertise to cover a global demand for knowledge utilisation in a specific domain.

The disruptive potential of these developments regarding the very structure of the way we cooperate, do business, and advance and use humankind's knowledge, necessitates a broad societal and technological discussion. Contributing to this discussion, we propose and explain key concepts of the stated vision, that are or will in our opinion be relevant for the described advancements.

The core elements of the envisioned socio-technical framework are the entities, that enable the scalable utilisation of knowledge, provided as automated service. For this purpose, the knowledge of an expert or group of experts ('owner') is represented by executable software ('code'), see Figure 1. The executables are made accessible by interfaces ('application programmable interfaces – APIs'). Via the API, certain software functions can be called to obtain specific information, or, in other words, to create answers to individual questions based on the represented knowledge. We call such an entity **Knowledge engine (KE)**, which is defined by these features: (i) It processes and generates (only) information (ii) automatically, (iii) based on represented knowledge – implicitly or explicitly represented. (iv) It enables users to utilise the represented knowledge without necessarily comprehending it. The users can be human or non-human (e.g. other KEs).

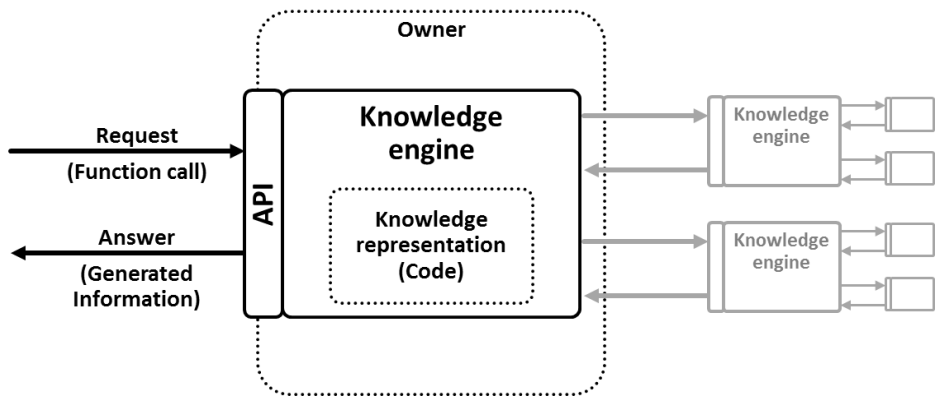

*Figure 1. The knowledge engine concept.*

Although advances in machine learning have gained most public attention recently, and significantly contribute to the creation of KEs, we expect versatile types of KEs to generate solutions in different domains. These will be built on very distinct technologies to represent and automatically apply knowledge, including:

(i) **Utilising explicit knowledge:** e.g. Simulations, answering the question: What are the properties of this given instance – based on these known laws (e.g. physical laws)? Knowledge-based systems, answering the question: What is the solution to this given problem – based on these known rules, experiment results, research studies, expert opinions, ontologies, etc.?

(ii) **Utilising implicit knowledge:** e.g. Supervised machine learning, answering the question: What are the properties of this given instance – based on these given instances and their known properties?

We state, that especially in domains, where fundamentally new knowledge and solutions are to be created (e.g. research, development, innovation), the importance of human reasoning and creativity, depicted by explicit knowledge is increased, as compared to the importance of machine-learned implicit knowledge, which has its strength in extracting the knowledge of past (human) decisions. The relation between the owner (one or multiple persons) and the KE is twofold: The owner creates, maintains and advances the knowledge that is represented within the KE. The owner can profit from business models accompanying the utilisation of the represented knowledge.

We call the set of all KEs, that are accessible via the internet, the **Internet of Knowledge engines (IoKE)**. This can be used to build more advanced problem-solving capabilities. In general, complex problems cannot be solved by one expertise alone. Knowledge from various sources needs to be combined. E.g. one cannot provide a KE to stabilise an electrical power grid without taking into account the knowledge-based service of a weather report, predicting the performance of wind turbines and solar farms, and electrical heating demand. Therefore, KEs will be linked to each other, whereas outputs of one KE serve as input for another one. In this way, complex automated workflows can be created. We call the set of all KEs involved in such a workflow **Network of Knowledge Engines (NeoKE)**. KEs within a NeoKE are also called sub-KEs. A NeoKE can be considered a (virtual) KE itself. We believe that at first, NeoKEs will reflect existing scientific and business relations, automating existing networks of value creations. But we expect them to create new forms of cooperation with completely different granularity of labour diversification and time scales. This might also include a fundamental cultural shift in how society relates to knowledge and to how knowledge creates value.

Within the concept of NeoKEs, it is possible to
- remunerate knowledge services, and to create an ecosystem of business cases and intrinsic motivation to let knowledge be utilised by others,

- maintain ownership and to not disclose intellectual property, as KEs can be operated as black boxes,
- access and combine knowledge without centralised aggregation,
- provide complexity reduction such that non-experts can utilise expert knowledge,
- transparently document the applied KEs.

### 2.4 Self-composing knowledge: Second-order automation

In the ongoing digital transformation, we are already increasingly relying on the benefits of connected services. One service can trigger a multitude of other services. Data, generated by one service, can be used to build others. By increasing the granularity of these services, the arising complexity is immense, due to technical, organisational, economical and legal challenges. These challenges create strong market forces towards centralised structures: Data ownership, proprietary standards, network effects, platform business models and economies of scale. The resulting market concentration undermines competition principles and – even more severe – endangers all types of innovation that originate outside these centralised structures. For instance, no innovative services depending on data and rights owned by dominant players can be realised without those players. The brute force antidote would be anti-trust laws. But there is also the option to create a balancing force towards a versatile and heterogeneous eco system: The reduction of transaction costs to enable the connection of distributed knowledge-based services.

In Section 2.3 the concept of KEs was introduced to depict accessible services to solve problems based on automated application of knowledge. This makes knowledge accessible and utilisable at very low marginal costs. However, these KEs must also be **findable, discoverable,**[8] **accessible and integrable** into larger workflows to solve superordinate tasks. On top of that, NeoKes can be considered as a type of reasoning. From this perspective, the generation of NeoKEs is a higher level of intelligence, creating this reasoning. The feasible granularity and flexibility of problem solving with NeoKEs is directly linked to the abilities to compose NeoKEs to solve specific problems. The effort of NeoKE composition relates to the concept of transactional costs.[9] In the same pace that marginal costs of knowledge utilisation itself are reduced by accessible automated services, the reduction of transactional costs will be in the focus of future advancements: The **self-composition of NeoKEs** is next. This quest can also be considered as second-order automation of knowledge work.

A sketch of how NeoKEs might be composed is shown in Figure 2. A human user chooses an access point, i.e. a KE with a user interface, to specify his problem or question. The access point can be rather generic, like a search tool with free text or speech input. Or it can be very specific, like a configurator to get an individually optimised product design. Composing KEs identify and call other KEs (sub-KEs), that are suitable to solve the defined problem. The composing KE creates and orchestrates the workflow between these sub-KEs. The composing KE is naturally embedded within NeoKEs.

Some examples for composing KEs are: (i) A rather simple composing KE could be a search tool: It calls a huge amount of sub-KEs, e.g. to find a requested piece of information. (ii) An example of a composing KE from the domain of modelling and simulation could be an optimiser that analyses properties of a virtual system by varying and testing the models of its components. For each variation, certain sub-KEs are called, each providing a specific model (or the black-box behaviour) of a component, to run a systemic co-simulation. (iii) A composing KE in machine learning could use received training data for a classification problem. It can analyse the structure of the data. Based on explicit rules it decides which types of classifier are likely to perform well. It calls sub-KEs of classifiers, provided by various research institutions, and compares the performance. It returns the optimal algorithm (e.g. trained neural network) along with a documentation on the reliability.

The set of sub-KEs that are involved in a NeoKE can be predefined manually (hard-wired) or selected automatically, e.g. at runtime, by the composing KE (self-composing NeoKEs).

In Section 5.2 we describe actual examples towards this process. The neighbouring domains of software libraries and data bases symbolise relevant ends of the spectrum of possible KEs. Knowledge-intensive software libraries with intuitive interfaces or data bases with advanced filtering and analytics tools are some examples of KEs from these domains.

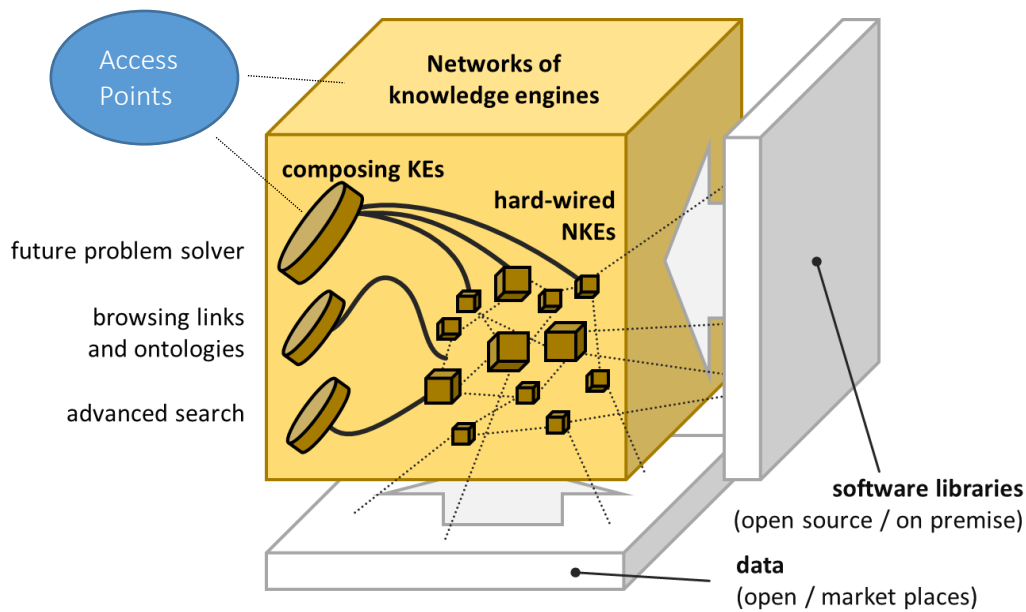

*Figure 2. Types of knowledge engines and network composition. Related concepts of software libraries and data market places.*

The self-composition of general web services (KEs are a specific type of web services) is an open research challenge. They are "far from reality. A more realistic approach is to provide context-aware recommendations in a composition's design phase."[10] However, we argue that KEs allow new approaches for self-composition, as integral knowledge representations (e.g. rules, models) can be utilised to identify suitable workflows for problem solving. E.g.: A KE, modelling the design of an electric drive, is prone to be connected to a KE, modelling a gear box. This composition is inspired by physical relations. We also argue along the notion of "findability being (also) a content problem, not a search problem",[11] to develop bottom-up approaches for self-composition.

## 2.5 ONGOING INITIATIVES

We highlight some initiatives which are directly contributing to realizing networks of knowledge engines. Of course, the basis of all of this is the trend of sharing data and knowledge, via wikis, open libraries, repositories, forums, open data deposits and service platforms. Additionally, initiatives structure knowledge, knowledge graphs, via manually created ontologies (e.g. Cyc[12]) or via ontology engineering or automation. The following concepts and initiatives deserve specific attention, as we consider them in close relation to the evolution of NeoKEs.

**Open source:** The concept is around for a long time: People are casting their knowledge into code and provide it to the rest of the world – constituting the open source movement. Within this approach the code itself is disclosed. Open-source has already substantially changed the co-creation of digital value. Likewise, **Linking/Open Data** Initiatives for freely available data (open data), for interlinking open data and research results are on the rise.[13, 14, 15, 16]

The **API-economy**. The API-economy arose from the concept of equipping fragments of software with web interfaces. Hence, these fragments can be composed to more complex software tools by external users. Provider of "APIs" can focus on performing specific tasks very well, without caring about the full tool chain (e.g. user interfaces). Software tools with APIs could be considered KEs, when they are based on substantial knowledge. Also, the related concepts of **service-oriented architecture** and **microservice architecture**, each having slightly different origins and views, provide useful groundwork for NeoKEs.[17] The automated composition of web-based services within the API economy is an open research question.[18,19] As we have argued herein before, in KEs the available knowledge representations can open new approaches to automated composition.

**Ontology web language for services** (OWL-S) is an OWL-based web services ontology, which provides the ability to describe the semantics of web services and their capabilities in a formal and machine-processable manner. Moreover, it aids semantic service matching, selection and composition,[20] with related approaches to learn from historic matching data.[21]

**Machine learning** got lots of attention in recent years, more specific: supervised deep learning (e.g. image recognition, speech recognition) and enforcement learning (e.g. alpha go). Machine learning can be conceived as getting implicit knowledge representations (e.g. a trained neural network) and sometimes also explicit knowledge insights ("explainable AI") from existing data, which already contains this knowledge (e.g. past human decisions, which the AI tries to mimic). The trained network or the process of training itself can be considered KEs.

**Inference systems and semantics reasoners** aim at providing logical consequences upon explicit knowledge and the derivation of ontologies. Initiatives and implementations along knowledge graphs and ontologies exist and are currently evolving such as OML reasoners, CYC or ontology engineering.

Within the **theory of work systems**, NeoKEs can be perceived as automated (at least regarding execution), inter-organisational information systems (devoted to the processing of information),[22] that are based on knowledge.

**Simulation and optimisation based on interacting models:** In engineering, many approaches exist to solve problems on a system level (e.g. design, configuration, control) by utilising interlinked virtual models of the constituting components to simulate and optimise the system behaviour. These approaches include model-bases system engineering (MBSE), concurrent engineering, knowledge-based engineering (KBE), virtual commissioning, interacting digital twins. When making such models from different providers available and connectable via platforms or system configuration tools, the vision of the IoKE and operational NeoKEs is already realised for these domains.

All described initiatives, trends and implementations show essential contributions towards an IoKE and NeoKEs, even self-composing NeoKEs, yielding promising ground work and synergies. These initiatives will increase the maturity and size of the IoKE and the demand for interconnectivity tools and framework elements, as addressed later on in Section 4.1. In Figure 3 we map several relevant trends and approaches with respect to two dimensions: The interaction range describes how many KEs a KE can potentially communicate with. This number is limited mainly by a lack of compatibility. The interaction type explains how generic or how specific the interaction is. A specific interaction can utilise all functionalities, hence the full depth of the knowledge provided by a KE. For example, a free-text search tool can communicate with any other KE, that is able to provide any natural plain text (function names, documentation). However, it does not interact on other levels than plain text.

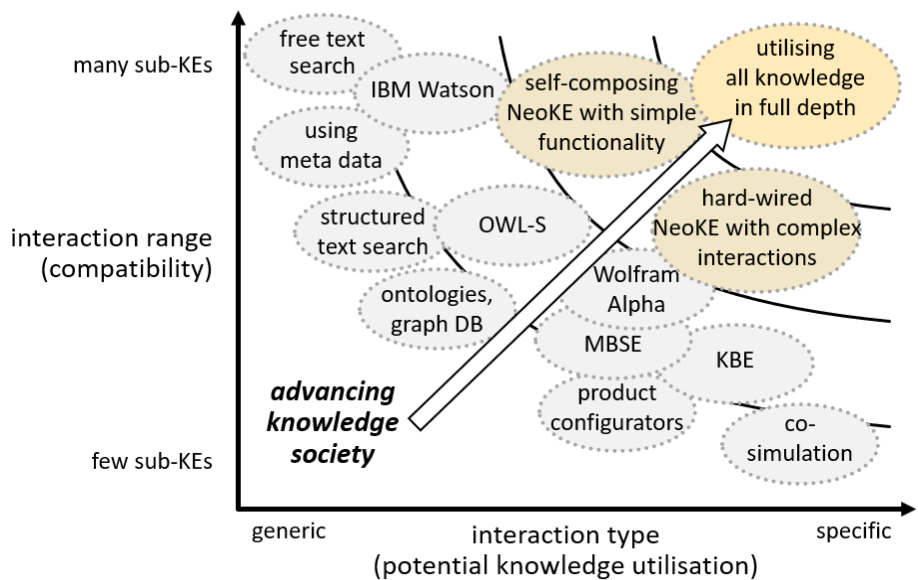

*Figure 3. Present and future concepts of knowledge utilisation, intuitively mapped with respect to two aspects: The interaction range describes with how many knowledge sources (sub-KEs) a given method can communicate with. The interaction type explains how generic or how specific the interaction can be.*

## 3 IMPACT

NeoKEs make knowledge digitally available, instantly deployable and not centrally governed. They enable a society that benefits from its collective knowledge in a decentralised manner with vivid business models incentivising experts to make their knowledge utilisable. This is expected to lead to unprecedented prosperity, when prosperity is understood as "the accumulation of solutions to human problems".[23] The realisation of NeoKEs, is expected to have the following implications on economy and society, some of which already rise within the evolution levels towards NeoKEs (see Section 5.1).

### 3.1 ECONOMY

Within the **digital single market**, customers search on platforms through product databases. Such search behaviour favours already existing, concrete products and penalises on-demand tailored solutions, as they are hardly depictable as a concrete product, ready for comparison. By deploying NeoKEs, a customer, with a formulated specific need, receives concrete solution options to this need. This concrete solution options can be off-the-shelf products or instantly composed (not yet existing) products (built by merging services and tailored abilities available via KEs from companies). The customer can now directly compare alternatives to satisfy the need. The deployment of NeoKEs facilitates the comparison of service-related solutions and products. This enables especially SMEs, whose assets are typically high flexibility, lot size one production and individualisation, to better compete with large corporates and their economies of scale.

**Leveraging the innovation potential.** Value creation is increasingly generated by interlinking domains and disciplines.[24] A nature of this interdisciplinarity is to enter fields the innovator is not an expert in. It takes substantial time and effort to identify the right players, competences and experts before interdisciplinary innovations can be established. NeoKEs reduce the exploration phase drastically, as self-composing entities bridge disciplines and connect the appropriate elements and dimensions. Doing so, NeoKEs **lower the barriers between disciplines** and leverages the take up of interdisciplinary innovation. Also, the instant deployment of expert knowledge via NeoKEs allows **fast iterations** and speeding up (design thinking) innovation actions.

**Modularizing knowledge.** Within a company, NeoKEs allow fast access to a company's digital assets and know-how. It enables knowledge management and preservation. With a proper auditability layer, it allows to identify how existing knowledge is contributing to a company's success. Across companies, NeoKEs allow the monetarisation of modularised knowledge. This allows new business models and essential revenue streams (also applicable as a return on public investments in research). The decentralised storage and black-box mentality of KEs allows proper handling of IPR, authorship and revenue.

## 3.2 SOCIETY

For **societal knowledge** management in libraries, museums and innovation clusters, the instant applicability of KEs transforms the storage functionality and accessibility of libraries into innovations centres, with instant usability, and ideation. Based on suitable auditability layers, NeoKEs can pinpoint heavy utilised knowledge areas and fruitful ecosystems. Those can be leveraged by strategic decisions of policy makers and the effectiveness of funding can be determined.

For the **general society** NeoKEs provide an empowerment to obtain tailored and trustworthy answers to individual user requests. This allows to counteract algorithmic infiltration and provides an expert system with low entrance barrier to users by reinforcing civil empowerment. Networks of knowledge engines empower individuals to investigate fake news and populist statements.

For a **transparency of inherent dynamics** and to reinsure **networks of trust**, NeoKEs deliver results on demand via an auditability layer. This traceability of knowledge creation is essential for understanding the network dynamics of knowledge use. For investors, it will become apparent which new results are used and which are not, allowing a better focus and evaluation of funding instruments. For academia it allows to better understand, which results serve as building blocks for others. The insights might cause a shift from impact assessment based on citations towards an impact assessment based on empowerment.

For fighting the declining **sense of purpose** (compare Marx's "Entfremdung") due to increased specialisation and division of labour in supply chains and services, NeoKEs reveal how a knowledge-intensive service contributes to prosperity generation – where prosperity is defined as *"the accumulation of solutions to human problems"*[29] (creating economic, societal or individual value).

# 4 CHALLENGES

To realise NeoKEs, the core functionalities need technical developments and an interconnectable framework. In order to create a flourishing eco system, also societal and legal dimensions need to be addressed to provide sustainability, credibility, liability and business cases.

## 4.1 TECHNICAL DEVELOPMENTS

**Increasing the pool of Knowledge Engines – forming an Internet of KEs.** Within the progressing automation of knowledge work,[25] more and more knowledge is represented by algorithms, which can be executed automatically and at low cost. To become instantly applicable for large user groups, an algorithm must be equipped with a standardised API. Together they constitute a KE, as described above. KEs must be accessible and findable constituting the internet of KEs (IoKE). The ability to operate KEs as black boxes on a server, held by their owner, allows to keep the knowledge save, where needed. This is a prerequisite for most business cases. A KE needs to be executable, either as a stand-alone or by utilising a referenced run-time environment. Tools and initiatives for such KEs already exist, see Section 2.5 and Section 5.2.

**Coordinating the formation of NeoKEs – self-composition.** To create information utilising various sources of knowledge, multiple KEs have to be connected to form a NeoKE. The forming process is in general expected to be of heavy complexity, see the "second order automation of knowledge work" described in Section 2.4. For an automated composition of NeoKEs, specific composing KEs need to be put in place. This might include advanced KE search tools and KE ontologies to navigate within the IoKE. Methods of current AI might be applicable. Another approach might be to add some sort of self-composing ability to KEs. For first steps towards a realisation, the connections could be established manually, see Section 5.

**Running NeoKEs.** For executing NeoKEs, we need execution tools, i.e. KEs that feed the appropriate input into called sub-KEs and coordinate data handover. In general, parameter variations are expected, and the execution tool will need to run multiple iterations, set success criteria and coordinate and terminate running executions. NeoKEs might even contain circular references between co-depending KEs, that need tools to resolve these relations at runtime (compare the domain of co-simulations). Execution tools have to manage trade-offs between non-disclosure of knowledge (facilitated by black boxes on remote private servers) and execution efficiency (facilitated by a centralised server). For versioning and auditability an execution within NeoKEs might need to result in: i) an answer to the problem statement and in ii) an auditable documentation, that could even allow to reproduce the execution at later points in time.

**Solution representation to humans.** The output of NeoKEs might not be human understandable. Outputs might need to be condensed, plotted and processed such that humans can make sense out of them. Tools/KEs are needed that use the computation results to form an intelligible answer to the initial user request. The output of a NeoKE might also be used feed real-world entities (e.g. devices) with automated expert-knowledge-based decisions.

**Access points to NeoKEs.** We need ways how a human user can formulate his need and get access to the network. Such an entrance gate could be a platform with users picking preformed elements and modelling their request according to a given structure. Or such an entrance gate could be a free text interface with a user describing his need by natural language. Of course, entrance gates could have any hybrid form, and several levels of user interaction until the need is sufficiently specified for processing. Such entrance gates form a spectrum: from no-code easy user interfaces to software development environments to create and evolve KEs throughout their lifecycle.

## 4.2 Societal and legal dimension

Beyond the challenge of creating the core functionality of NeoKEs, we have identified several pressing societal and legal aspects related to NeoKEs:

**Certificates.** NeoKEs will be storing knowledge and will return answers to users. In order to give meaningful answers, a quality management system has to be installed. As KEs themselves, belong to its creator and owner and are often black boxes to users, the proper functioning and the proper knowledge derivation need to be certified.

**Auditability.** The ability of a NeoKE to document how it derived a result is essential. Objectivity is a key element for reasoning and credibility. All user interfaces need to be able to trace back to every single element that was needed for a response to each single user request. It might be possible that due to different access points to NeoKEs different KEs are deployed and derive different results. It is essential to be able to trace back and reason these outcomes (e.g. for credibility issues). Current technology developments such as blockchain might provide the right ledgers for documentation and credibility creation of NeoKEs.

**Verification units and versioning.** KEs are software, which is prone to bugs and will need to experience repeated updates. Procedures would need to be installed to check the system is not crashing if new KEs are installed. While NeoKEs are conceptualised as self-organizing and decentral networks, it might be necessary to have centralised guards, which check for emerging unexpected dynamics. To comply with both auditability and versioning, it might be necessary to constantly back-up NeoKEs in order to respond to law suits and to cover the liability concerns.

**Global constituency – who enforces it?** KEs are supposed to provide knowledge and meaningful results. While a single KE might have its certificate and the creation of a NeoKE can be tracked, it is not obvious that a NeoKE would provide meaningful, yet correct results. Also, the ability to form NeoKEs might need the existence of local, super-local or global ontologies. The creation of such ontologies might need a human (?) supervision body and a constituency. Appropriate forms and processes need to be developed to derive such a constituency. As soon as NeoKEs are able to generate new knowledge engines, the world automatically moves closer to technological singularities. The global constituency and supervision also need to keep in mind how to handle and design an internet of knowledge engines and its designed limitations.

**Liability on KEs.** The KEs are the essence of the knowledge system. The contribution of such engines comes by different motivation, to make money (see below), to contribute to a vision and to comply with the law (below). In all cases, different liability restrictions apply, (just as with legal entities in general). To integrate liability in NeoKEs, KEs could for example specify the limits of liability, and to return this along with NeoKEs' results. NeoKEs that combine different KE input would also need to interpret the different liabilities. New form of licencing could emerge, comparable to those in open source software. In any case, the liability of KEs will occupy the attention of lawyers, and countries could think of the installation of public KE licencing. This can help companies in defining their liabilities with KEs, and/or to take over liability issues from the owner for certain fees. Doing so, the state could create a new source of public income, while fostering knowledge-based value creation.

**Differentiation of labour and centralisation of expertise.** KEs provide the possibility of scalable knowledge utilisation as marginal costs are low. Maybe even more crucial, a KE that provides services for many users, would also more likely get feedback from those users. (For many services a direct data feedback would in fact be necessary.) Hence, a KE with a large user base could operate at lower average costs and learn faster than a KE with a small user base. This leads to centralisation of the expertise in a specific field of interest. It might even be sufficient and economically favoured to have a single solution provider for each problem niche. Political strategies need to be put in place to allow or counteract monopolistic KEs.

## 4.3 BUSINESS CASES

While effortless copying due to digitisation threatened quite some industries (e.g. music industry), some initiatives such as Wikipedia rely on donation to keep their digital business going. With NeoKEs, it becomes possible to let others utilise one's knowledge, without necessarily disclosing the knowledge. This paves the way for new kinds of IP-based business models beyond patents and associated licensing.

**New business models.** In the world of KEs, the knowledge is within the engine and not copiable, as music for example. This allows to experiment with new business models such as a freemium model. At the beginning the user wants to get a "fast feel" upon the request. If that is enough for the user, the relevant KE would need to consider the costs as advertisement. But if the user is interested in specifications, detailed documentation or the assumption regarding liabilities, the user might be more likely to use fee-based services. Regarding the single digital market, the KEs can be used to make service related solutions and products comparable with off-the-shelf products (see Section 3.1).

**New Market dynamics.** In its techno-economical optimum a globally pervasive NeoKE will yield extreme scalability of knowledge utilisation and transparency regarding the performance of various KEs being active in the same domain. These are advantageous properties. But are society and its state of solidarity prepared to harness them?

## 5 Towards a Realisation

While most aspects of the proposed system still need heavy research for general applicability, we outline a technical evolution and also identify existing solutions with specific scope.

### 5.1 Evolution levels

To realise the overall vision, technological evolutions needs to take place with each application area and across disciplines. The implementation dept varies from case to case. The classification from Figure 4 depicts which steps we expect to happen towards a realisation. As a base line, knowledge is uses manually. The first step is to derive a digital representation of the knowledge in the form of a knowledge engine. This can partly be performed via digitisation or machine learning. Second, is the collective availability of the KEs on internal or open platforms. As a third step, the KEs are systemised and can be browsed within a logic that represents dependencies, links, possible extensions, and usage patterns. Final the forth steps is to install composing KEs that that are able to form NeoKEs upon individual requests.

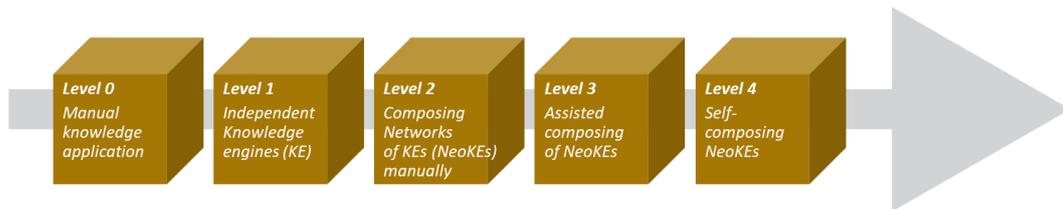

*Figure 4. The evolution of Networks of Knowledge Engines. To reach full capability several technological levels will need to be passed.*

### 5.2 Implementation examples

Out of the pool of initiatives we pick two examples at current frontiers towards a realisation of NeoKes.

**Wolfram Alpha**. "is a unique engine for computing answers and providing knowledge"[33] and is a question answering tool. It is operated and owned by Wolfram Research. [26] It allows natural language problem statement and returns a visually enriched answer. For answer generation Wolfram Alpha uses so called "computational knowledge engines". These are discipline specific simulations and knowledge representations used to create responses. The tool is a great example of how KEs are integrated and applied in specific disciplines for answer derivation. Towards the realisation of NeoKes, within our logic, the next step is to establish the ability of combining KEs, such that the output of one KE is used as an input by another KE. This leads us to our second example.

**SyMSpace**[27] is a simulation platform to optimise design processes. It is operated and owned by Linz Center of Mechatronics GmbH. It originated from its primary use case which is for automating the design process of electric drives, where it reduces development time from months to days. The platform serves as an execution pipeline and knowledge space. It handles various KEs, which can be linked by drag and drop. The resulted chain of KEs is executed and attached to an optimiser to derive the quasi best parameter sets. The platform is open source and offers templates to attach further KEs, which are utilised by its industry customers. Due to its generic system design, SyMSpace it currently piloted for discipline extension and business case exploration as envisioned in Section 3 and 4.